\documentclass{aa}
\usepackage{txfonts}
\usepackage{rotating}
\usepackage{answers}
\usepackage{xcolor}
\usepackage{graphicx}
\usepackage{scalefnt}
\usepackage[colorlinks=true,allcolors=blue]{hyperref}
\usepackage{adjustbox}
\usepackage{multirow}
\usepackage{tablefootnote}
\usepackage[rightcaption]{sidecap}
\defcitealias{harris10}{H10}
\defcitealias{perez20}{PV20}
\makeatletter
\renewcommand*\aa@pageof{, page \thepage{} of \pageref*{LastPage}}
\newbox\grsign \setbox\grsign=\hbox{$>$} \newdimen\grdimen \grdimen=\ht\grsign
\newbox\simlessbox \newbox\simgreatbox
\setbox\simgreatbox=\hbox{\raise.5ex\hbox{$>$}\llap
     {\lower.5ex\hbox{$\sim$}}}\ht1=\grdimen\dp1=0pt
\setbox\simlessbox=\hbox{\raise.5ex\hbox{$<$}\llap
     {\lower.5ex\hbox{$\sim$}}}\ht2=\grdimen\dp2=0pt

\def\simless{\mathrel{\copy\simlessbox}}
\begin{document}

\title{A census of new globular clusters in the Galactic bulge}
\author{
  E. Bica\inst{1}
  \and
  S. Ortolani\inst{2,3,4}
  \and
B. Barbuy\inst{5}
\and
R.~A.~P. Oliveira\inst{6}
}
\offprints{B. Barbuy}
\institute{
Universidade Federal do Rio Grande do Sul, Departamento de Astronomia,
CP 15051, Porto Alegre 91501-970, Brazil
\and
Universit\`a di Padova, Dipartimento di Astronomia, Vicolo
dell'Osservatorio 2, I-35122 Padova, Italy
\and
INAF-Osservatorio Astronomico di Padova, Vicolo dell'Osservatorio 5,
I-35122 Padova, Italy
\and
Centro di Ateneo di Studi e Attivit\`a Spaziali “Giuseppe Colombo” – CISAS, Via Venezia 15, 35131 Padova, Italy
\and
  Universidade de S\~ao Paulo, IAG, Rua do Mat\~ao 1226,
Cidade Universit\'aria, S\~ao Paulo 05508-900, Brazil
\and
 Astronomical Observatory, University of Warsaw, Al. Ujazdowskie 4, 00-478 Warszawa, Poland
}
\date{Received; accepted }
 \abstract
     {The number of known globular clusters in the Galactic bulge has been increasing steadily
     thanks to different new surveys.}
   {The aim of this study is to provide a census of the newly revealed globular clusters in the Galactic bulge,
   and analyze their characteristics.
}
   {In recent years, many globular clusters have been discovered or identified. The stellar populations to which they belong are 
   indicated in their original studies: they are mostly bulge clusters,
   with some identified as disk or halo members. We collected 41 new globular clusters revealed in the last decade and compared them to the known bulge clusters. }
   {The new clusters are intrinsically faint with $M_{\rm V}$ of around $-6.0$\,mag.
The distance to the Sun of the ensemble of well-known and new bulge clusters is compatible
 with the Galactocentric distance measurements from the Galactic black 
hole location.
{The ensemble sample} shows metallicity peaks
at
$\rm{[Fe/H]}\sim -1.08\pm0.35$ and $-0.51\pm0.25$\,dex,
 confirming previous findings.
The age-metallicity relation of the new clusters younger than 10\,Gyr is compatible with that of the ex situ samples of the dwarf galaxies Sagittarius, Canis Majoris, and Gaia-Enceladus-Sausage. 
The clusters with ages between 11.5 and 13.5\,Gyr
show no age-metallicity relation, because they are all old.
{This is compatible with their formation in situ in the early Galaxy.}
}
   {}
\titlerunning{A census of new globular clusters in the Galactic bulge}
\authorrunning{E. Bica et al.}
\maketitle
%

\section{Introduction}

The Galactic bulge probably formed due to early mergers within the $\Lambda$CDM
scenario, and its present configuration was completed later with a buckling bar
\citep[e.g.][see also review by \citealt{barbuy18a}]{queiroz20, queiroz21}.

The origin of globular clusters (GCs) in the Galactic bulge has become better
understood in recent years, thanks to the proper motion measurements from the
\citet[and references therein]{gaia23}, which allow one to examine their orbits.
The nature of the GCs in the Galaxy having two branches in the age-metallicity relation (AMR) was revealed through
the derivation of relative ages from ACS/\textit{HST} data by \citet{marin-franch09}. 
The age-metallicity bifurcation using their data is clearly seen in figure~10 from \citet{barbuy09},
for example, which shows one branch with the old clusters, in principle formed in situ in the early Galactic bulge, and another branch with the younger ones.
This scenario was completed by \citet{forbes10}, making evident that the AMR of the younger
clusters fits well that of the dwarf galaxies Sagittarius and Canis Majoris, indicating that they should be accreted ex situ clusters,
whereas the older clusters are all old and were formed early on in the history of our Galaxy.
According to \citet{forbes20}, 73 of the known GCs were formed in situ, very early in
the Galaxy's history, and 87 were accreted from dwarf galaxies that merged
with the Galaxy. Similar classifications were proposed
by \citet{massari19}, \citet{kruijssen20},
\citet[hereafter \citetalias{perez20}]{perez20},
\citet{callingham22}, \citet{horta23}, and \citet{belokurov24}. These authors show the contrast between the age-metallicity plot of
in situ clusters, and that of dwarf galaxies.

\citet{bica16} identified 43 well-known GCs that would belong to
the Galactic bulge, based on an angular distance to the Galactic center lower than 20$^{\circ}$, a distance to the Galactic center lower than $3.5$\,kpc,
and a metallicity of $\rm{[Fe/H]}>-1.5$\,dex. It is interesting to point out the
consideration of metallicity by
\citet{geisler23},
suggesting that clusters with
metallicities lower than $\rm{[Fe/H]}\simless-1.5$\,dex should be identified as more probably being halo clusters.  
Recently, \citet{horta23} and \citet{belokurov24} classified clusters
as in situ or ex situ using ages, metallicities, detailed chemistry, total energy, and the $z$ component
of the orbital angular momentum.

The sample of Milky Way GCs is still incomplete, as was pointed out, for example, by \citet{ivanov05}, \citet{kurtev08}, and \citet{bica19}, and progress in the detection of newly identified GCs has been made in recent years.
Most of the newly revealed clusters were detected in the last five years,
due to a combination of the VVV \citep{minniti10} and VVVX surveys \citep{borissova18} in $J$, $H$, $K_{S}$, and
proper motions from \textit{Gaia} Data Release 3 \citep[DR3;][]{gaia23}, which allowed
decontamination from field stars.

The present work deals with the faint and in general low-mass
clusters revealed by several surveys, such as VVV/VVVX  in the near-infrared
and WISE \citep{wright10} in the mid-infrared.
A number of new confirmed and candidate GCs
are now in the process of being identified or having their
existences confirmed, and metallicity, reddening, age, distance, and kinematical information is being derived. 
\citet{bica19} present a catalog
of Galactic clusters and associations, and compile 200 GCs, plus 94 GC candidates.
\citet{baumgardt21} and \citet{vasiliev21} report a list of known GCs,
and derived their distances by making use of {\textit{Gaia} DR3} \citep{gaia23} proper motions, including some
objects in common with the list presented in this paper, as is described below.

In order to study these objects, we compiled the literature data for
39 confirmed clusters and two GC candidates, with 19 of them located in the bulge, considerably increasing the previous number of known bulge clusters \citep{bica16}.

A compilation of newly revealed clusters is described  in Sect.~\ref{sec:2}. Previously identified bulge
clusters, with updated data, are reported in  Sect.~\ref{sec:3}.
Ages, distances, and metallicities are analyzed in  Sect.~\ref{sec:4}.  Conclusions are
drawn in Sect.~\ref{sec:5}.

\section{New globular clusters identified in the Galactic bulge}
\label{sec:2}

In Table \ref{logbook} we report 39 newly identified GCs and two GC candidates, basically since the last release of the \citet[2010 edition\footnote{\url{https://physics.mcmaster.ca/~harris/mwgc.dat}}, hereafter \citetalias{harris10}]{harris10} catalog,
and in particular since the \citet{bica16} review.
The table lists coordinates, reddening, distance, metallicity,
age, proper motions, absolute magnitudes, and an indication of the stellar
populations to which they belong. Most of them can be identified as bulge
clusters on the basis of coordinates and distance.

Below, we provide a few comments on some individual clusters. Among the new clusters by \citet{camargo18} and \citet{camargo19}, 
for which proper motions are not explicitly given in the papers, but
rather shown in plots, 
Camargo 1102, 1103, and 1104 have a different proper motion relative to the field;
Camargo 1107, 1108, and 1109, on the other hand, have an indistinguishable proper motion, but
deep VVV color-magnitude diagrams (CMDs), which suggests that they are real clusters. 
{We note, however, that \citet{gran19} discard these clusters and dozens of Minniti clusters based on \textit{Gaia} DR2 and VVV photometry and proper motions, validating only VVV-CL001 \citep{minniti11} and VVV-CL002 \citep{moni-bidin11}. \citet{gran19} state, however, that for these candidates to be real they must have low mass and/or low concentration, or proper motions similar to the field stars (which happens frequently up to 5\,mas/yr in their appendix figures). We prefer to move only Camargo 1105 and 1106 to the bottom of Table~\ref{logbook} as candidates, and wait for spectroscopy and deeper photometry before definitely confirming or discarding the other Camargo candidates.}

\citet{horta23} and \citet{belokurov24} have presented evidence of clusters
having formed in situ or having been accreted. Most of the new clusters are not included in the studies,
but one case is of particular interest: FSR 1758. \citet{myeong19} and \citet{vasiliev21}
have shown, based on its retrograde orbit and other properties, that FSR 1758 was accreted from the Sequoia dwarf galaxy.

\citet{gran22} report radial velocities and orbit calculations and conclude that Gran 1 is a bulge GC, formed in situ and that Gran 2, 3, and 5  might have their origin in the Gaia-Enceladus-Sausage structure. 
\citet{belokurov24} instead classify the latter clusters
as having formed in situ.

It is also important to point out the measurement of
radial velocities with the IGRINS/Gemini spectrograph by \citet{garro23}, and the calculation of orbits.
They find that 
Patchick 125 and Patchick 126 are bulge (or halo) clusters, that Ferrero 54, Gaia 2, and Patchick 122
should be disk clusters, and that VVV-CL160 is close to the
Galactic center but orbits beyond the solar circle, and therefore has an unclear origin.

Finally, we note that the new clusters that lie outside the range of distances studied here,
the ones that have metallicities lower than $\rm{[Fe/H]} \leq -1.5$\,dex, 
and those classified as halo or disk members,
are excluded from the plots showing the location of clusters in galactic coordinates, as well as from 
the metallicity and distance distributions. This amounts
to 22 clusters that are not considered; that is, 19 new clusters are included as bulge members. 

\begin{table*}
  \caption{Literature data on the 41 newly identified GCs and candidates, reporting coordinates, reddening, distance to the Sun, metallicity, age, proper motions, magnitudes, and assignation to a Galactic stellar population. }
\scalefont{0.7}
\begin{flushleft}
  \begin{tabular}{l@{}cccccrrrc@{}cr@{}r@{}r@{}r@{}c@{}c@{}c@{}c@{}c@{}}
\noalign{\smallskip}
\hline
\noalign{\smallskip}
\hline
\noalign{\smallskip}
Name & RA  &  Dec. & $\ell$ & $b$ &  A$_{\rm Ks}$ &  d$_{\odot}$ & [Fe/H] & age &  $\mu_{\alpha}^{*}$ &  $\mu_\delta$  &  M$_{\rm K_{s}}$ &    M$_{\rm V}$ & Pop. &  Ref. \\
 & ($^{\circ}$)  &  ($^{\circ}$) & ($^{\circ}$) & ($^{\circ}$) &  & (kpc) & (dex) & (Gyr) &  \multicolumn{2}{c}{\hbox{(mas/yr)}}   & \multicolumn{2}{r}{\hbox{(mag)}}  &  &  \\
\noalign{\smallskip}
\hline
\noalign{\smallskip}
Gaia 2        & $28.138$  & $53.043$ 
& 132.155 & $-$8.730 & 0.10 &   4.9$\pm 0.5$ & $-0.90\pm 0.2$ &  10.0$\pm 1$   & $-1.31\pm0.18$ &    $1.21\pm0.19$ &
$-5.4\pm2.0$  & $-$3.9  & D & 1 \\ 
Ferrero 54    &  $128.451$ & $-44.447$ 
& 262.803 & $-$2.570 & 0.60 &    7.1$\pm 0.4$ &  $-0.2\pm 0.2$ & $>$10 & $-1.33\pm0.27$ & \phantom{-}$1.31\pm0.34$ & \phantom{-}$-5.87\pm1.7$ & \phantom{-} $-$4.1 &  D & 1 \\ 
Patchick 122  &  $145.628$ & $-52.428$ 
& 276.340 &    0.405 & 0.63 &    5.6$\pm 0.4$ &  $-0.5\pm 0.2$ &  $>$6 & $-3.72\pm0.12$ & \phantom{-}$3.81\pm0.12$ & \phantom{-}$-6.12\pm1.0$ & \phantom{-} $-$4.2 &  D & 1 \\ 
Garro 1       & $212.250$ & $-65.620$ 
& 310.828 & $-$3.944 & 0.15 &      15.5$\pm 1.0$ &  $-0.7\pm 0.2$ &  11.0$\pm 1$ & $-4.68\pm0.47$ & \phantom{-}$-1.34\pm0.45$ & \phantom{-}$-7.76\pm0.5$ & \phantom{-}$-$5.26 &  D & 2 \\ 
RLGC 1         & $244.285$  & $-44.594$ 
& 336.870 &    4.303 &   --   &     28.8$\pm 4.3$ &  $-2.2\pm 0.2$ &  12.6 &   1.04$\pm0.05$    &   $0.83\pm0.04$    &    --   &  --     &  H & 3 \\ 
Patchick 125,Gran 3$^{I}$ &\phantom{-}$256.253$ & $-35.495$ 
& 349.756 &    3.424 & 0.33 &  10.9$\pm 0.5$ &  $-1.63\pm 0.14$ &  14   & $-3.85\pm0.50$ &    \phantom{-}$0.64\pm0.39$ & \phantom{-}$-6.1\pm0.8$  &  \phantom{-}$-$3.8 & B/H & 1,4 \\ 
Patchick 126$^{I}$  & $256.411$ & $-47.342$ 
& 340.380 & $-$3.825 & 0.44 &  8.6$\pm 0.4$ &  $-0.7\pm 0.3$ &  $>$8 & $-4.75\pm0.46$ & \phantom{-}$-6.68\pm0.62$ & \phantom{-}$-5.56\pm0.8$ &  \phantom{-}$-$3.5 & B/H & 1 \\ 
Gran 2$^{I}$        & $257.890$ & $-24.849$ 
& 359.229 &    8.587 &  --  &  16.6 & $-1.46\pm 0.1$ &  --   &    0.19 & $-$2.57 &    --   & $-$5.92 &  H & 4 \\ 
Camargo 1102  & $260.437$ & $-26.544$ 
& 359.145 &    5.734 &  --  &     8.2$\pm 1.2$ &  $-1.7\pm 0.2$ &  13.3$\pm 1$ &     --  &   --    &    --   &  
\phantom{-}$-6.3\pm 0.6$ & B & 5 \\ 
FSR1758$^{A}$       & $262.800$ & $-39.808$ 
& 349.217 & $-$3.292 &  --  &  11.5$\pm 1.0$ & $-1.5\pm 0.3$   &  --   & $-$2.19 &    2.55 &    --   &    --   & B & 6 \\  

Minni 48      & $263.325$ & $-28.001$ 
& 359.351 &    2.790 & 0.45 &   8.4$\pm 1.0$ & $-0.2\pm 0.3$   &  10$\pm 2$   & $-3.1\pm0.4$ & \phantom{-}$-6.0\pm0.5$  & \phantom{-}$-9.04\pm 0.66$ & \phantom{-}$-$6.5  & B & 7 \\ 
FSR019        & $263.910$ & $-21.070$ 
&   5.499 &    6.071 & 0.19 &       7.2$\pm 0.7$ & $-0.5$  &  11   & $-2.50\pm0.76$ & \phantom{-}$-5.02\pm0.47$ & \phantom{-}$-$7.72 & \phantom{-}$-$4.62 & B & 8 \\ 
FSR1767       & $263.929$ & $-36.358$ 
& 352.601 & $-$2.166 & 0.28 &      10.6$\pm 0.2$ & $-0.7\pm 0.2$  &  11$\pm 2$   & $-3.02\pm0.50$ & \phantom{-}$-4.85\pm0.50$ & \phantom{-}$-8.4\pm 1.5$  & \phantom{-}$-$6.3  & B & 9 \\ 

Camargo 1107  & $264.243$ & $-30.147$ 
& 357.977 &    0.956 & --   &       4.0$\pm 0.7$  & $-2.2\pm 0.4$   &  13.5$\pm 2$ &  --     &   --    &   --    & 
\phantom{-}$-6.6\pm 0.5$  & B & 10 \\ 
ESO393-12     & $264.657$ & $-35.651$ 
& 353.514 & $-$2.284 & 0.23 &   8.2$\pm 0.4$ & $-0.6\pm 0.2$  &  10$\pm 2$   & $-2.86\pm0.47$ & $-5.39\pm0.44$ & 
$-7.7\pm 1.5$  & $-$5.3  & B & 9 \\ 

VVV-CL003     & $264.728$ & $-29.907$ 
& 358.405 &    0.730 & 0.92 &     13.2$\pm 0.8$ & $-0.1$  &   --  & $-1.93\pm0.05$ &    $8.33\pm0.05$ & $-$9.92 & $-$6.82 & H & 11,12 \\ 

VVV-CL002     & $265.276$ & $-28.845$ 
& 359.559 &    0.889 & 1.07 &      8.6$\pm 0.6$ & $-0.4$  & $>$6.5& $-9.33\pm0.07$ &    $2.78\pm0.07$ & $-$7.71 & $-$4.61 & B & 11,12 \\ 

VVV-CL131     & $265.321$ & $-34.567$ 
& 354.721 & $-$2.170 & 0.23 &   9.0$\pm 0.5$ & $-0.6\pm 0.2$  &  10$\pm 3$   & $-3.24\pm0.81$ & $-5.65\pm0.07$ & 
$-8.2\pm 1.5$  & $-$5.9  & B & 9 \\ 
FSR025        & $265.430$ & $-19.571$ 
&   7.534 &    5.649 & 0.27 &      7.0$\pm 0.6$ & $-0.5$  &  11   & $-2.61\pm1.27$ & $-5.23\pm0.74$ & $-$7.31 & $-$4.21 & B & 8 \\ 
VVV-CL143     & $266.150$ & $-33.738$ 
& 355.788 & $-$2.319 & 0.21 &   8.9$\pm 0.5$ & $-0.6\pm 0.2$  &  10$\pm 3$   & $-3.18\pm0.91$ & $-6.17\pm0.85$ & 
$-8.2\pm 1.3$  & $-$5.9  & B & 9 \\ 
Camargo 1108  & $266.518$ & $-30.865$ 
& 358.404 & $-$1.087 &  --  &     3.3$\pm 0.5$  & $-1.8\pm 0.3$   &  13.5$\pm 1.5$ &   --    &  --   & --      & 
\phantom{-}$-8.4\pm 0.5$  & B/H & 10 \\ 
Camargo 1109  & $266.861$ & $-26.648$ 
&   2.165 &    0.844 &  --  &     4.3$\pm 0.6$  & $-1.5\pm 0.2$   &  12.0$\pm 1.5$ &   --    &  --     &   --    & 
\phantom{-}$-6.4\pm 0.7$  & B & 10 \\ 
Minni 22      & $267.214$ & $-33.061$ 
& 356.828 & $-$2.729 & 0.47 &      7.4$\pm 0.3$ &  $-1.3\pm 0.3$ &  11.2 &  --     &   --    &    --   &  $-$6.2 & B & 13 \\ 
Gran 5$^{I}$        & $267.228$ & $-24.170$ 
&   4.459 &    1.838 & 0.43 &  4.47 & $-1.02\pm 0.11$ &  --   & $-$5.32 & $-$9.20 &  --     & $-$5.95 & B & 4 \\ 
ESO456-09 & $268.476$ & $-32.466$ 
& 357.882 & $-$3.339 & 0.18 &   7.6$\pm 0.4$ & $-0.6\pm 0.2$  &  10$\pm 2$   & $-3.41\pm0.71$ & $-4.36\pm0.75$ & 
$-8.3\pm 1.5$  & $-$6.0  & B & 9 \\ 
FSR1776       & $268.558$ & $-36.152$ 
& 354.720 & $-$5.249 &  --  &      7.2$\pm 0.5$ & $0.02\pm 0.14$  &  10$\pm 1
$   & $-1.9\pm0.9$  & $-2.6\pm0.8$  &  --     &  --     & B & 14 \\ 
VVV-CL001     & $268.677$ & $-24.015$ 
&   5.268 &    0.780 &  --  &      8.2$^{+1.84}_{-1.93}$ & $-2.45\pm 0.24$ &  11.9$^{+3.12}_{-4.05}$ & $-3.41\pm0.50$ & $-1.97\pm0.50$ &   --    &  --     & HI & 15,16 \\ 
FSR1775       & $269.022$ & $-36.566$ 
& 354.546 & $-$5.779 & 0.16 &   8.9$\pm 0.2$ & $-1.1\pm 0.2$  &  10$\pm 2$   & $-3.00\pm0.80$ & $-5.53\pm0.73$ & 
$-8.0\pm1.7$  & $-$5.6  & B & 9 \\ 
ESO456-29,Gran 1$^{I}$ & $269.651$ & $-32.020$ 
& 358.767 & $-$3.977 & 0.24 &   7.9 & $-1.13\pm 0.06$ &  --   & $-$8.10 & $-$8.01 &  --     & $-$5.46 & B & 4 \\ 
Camargo 1104  & $271.309$ & $-24.979$ 
&   5.621 & $-$1.778 & --   &      5.4$\pm 1.0$ & $-1.8\pm 0.3$  &  13.5$\pm 0.5$ &  --     &  --     &   --    & 
\phantom{-}$-5.7\pm 1.7$  & B & 5 \\ 
Garro 2       & $271.463$ & $-17.701$ 
&  12.042 &    1.656 & 0.79 &   5.6$\pm 0.8$  & $-1.30\pm 0.2$ &  12$\pm 2$   & $-6.07\pm0.62$ & $-6.15\pm0.75 $ & 
$-7.52\pm 1.23$ & $-$5.44 & H & 17 \\ 
Camargo 1103  & $271.631$ & $-25.162$ 
&   5.604 & $-$2.121 & --   &      5.0$\pm 0.8$ & $-1.8\pm 0.3$  &  13.5$\pm 1$ &  --     &  --     &  --     & 
\phantom{-}$-6.9\pm 1.0$  & B & 5 \\ 

VVV-CL160     & $271.738$ & $-20.011$ 
&  10.151 &    0.302 & 1.71 &    4.0$\pm 0.5$ & $-1.4\pm 0.2$  &  13$\pm 2$   & $-2.90\pm1.28$ & $-16.47\pm1.31$ &
$-7.9\pm 1.5$ & $-$5.5  & B & 18,9 \\ 

Kronberger 49 & $272.600$ & $-23.340$ 
&   7.627 & $-$2.012 & 0.30 &   8.3$\pm 0.5$  & $-0.2\pm 0.2$  &  11$\pm 2$   & $-2.84\pm0.69$ & $-5.52\pm0.71$ & 
$-8.5\pm 1.5$ & $-$6.7  & B & 19,9 \\ 
Patchick 99   & $273.946$ & $-29.813$ 
&   2.488 & $-$6.145 & 0.09 &   6.4$\pm 0.2$ & $-0.2\pm 0.2$  &  10$\pm 2$   & $-2.59\pm1.51$ & $-5.49\pm2.02$ & 
$-7.0\pm 0.6$ & $-$5.2  & B & 20 \\ 
FSR0009       & $277.128$ & $-31.907$ 
&   1.855 & $-$9.529 & 0.11 &   6.9$\pm 0.2$ & $-1.2\pm 0.3$  &  11$\pm 2$   & $-1.39\pm1.10$ & $-5.22\pm0.99$ & 
$-5.8\pm 0.7$  & $-$3.4  & B & 9 \\ 
Gran 4        & $278.113$ & $-23.101$ 
&  10.198 & $-$6.388 & 0.14 &  22.5 & $-1.72\pm 0.32$  &  --   &    0.46 & $-$3.49 &  --     & $-$6.45 & H & 4 \\ 
RLGC 2         & $281.368$ & $-5.193$ 
&  27.631 & $-$1.042 & --   &     15.8$\pm 2.4$ & $-2.1\pm 0.3$  & 12.6  &  $-2.41\pm0.05$     &    $-1.85\pm0.05$   &  --     &         & B & 3 \\ 
Riddle 15     & $287.787$ & $14.833$ 
&  48.355 &    2.455 & 0.52 &  18.1$\pm 0.5$ & $-1.4\pm 0.2$  & $>$10 & $-1.03\pm0.32$ & $-1.64\pm0.27$ & 
$-7.6\pm 0.8$  & $-$6.2  & H & 1 \\ 
\noalign{\smallskip}
\hline
\noalign{\smallskip}
\multicolumn{15}{c}{  Globular cluster candidates} \\
\noalign{\smallskip}
\hline
\noalign{\smallskip}
Camargo 1106  & $263.143$ & $-30.280$ 
& 357.351 &    1.683 &  --  &      4.5$\pm 0.4$ & $-1.5\pm 0.3$  &  12.5$\pm 1$ &    --   &    --   &    --   & 
\phantom{-}$-5.7\pm 1.6$  & B & 5 \\ 
Camargo 1105  & $264.141$ & $-28.311$ 
& 359.479 &    2.017 & --   &       5.8$\pm 0.9$ & $-1.5\pm 0.2$  &  13.0$\pm 1$ &  --     &   --    &  --     &
\phantom{-}$-6.3\pm 1.0$  & B & 5 \\ 
\noalign{\smallskip}
\hline
\noalign{\smallskip}
\hline
\end{tabular}
\end{flushleft}
References:
    1 \citet{garro22b}; 2 \citet{garro20}; 3 \citet{ryu18};
    4 \citet{gran22} and \citet{gran24}; 5 \citet{camargo18};
    6 \citet{barba19}; 
    7 \citet{minniti21c}; 8 \citet{obasi21}; 9 \citet{garro22a};
    10 \citet{camargo19};
    11 \citet{moni-bidin11}; 12 \citet{minniti21a}; 13 \citet{minniti18}; 14 \citet{dias22};
    15 \citet{minniti11}, 16 \citet{fernandeztrincado21a}; 17 \citet{garro22c};
    18 \citet{minniti21b}, 19 \citet{ortolani12}; 20 \citet{garro21}. 
    $^{I}$ or $^{A}$ identified as in situ or accreted by \citet{belokurov24}.
    Population: B:bulge, D:Disk, H:Halo, HI:Halo intruder. Errors are adopted from the original papers, when available.
\label{logbook}
\end{table*}

The reported data were retrieved from the individual papers with the given references. For five of them, distances are also provided, based on \textit{Gaia} data, by
\citet{vasiliev21}.

Radial velocities are only available for a few clusters, reported in Table \ref{vr}.

\begin{table}
\caption[4]{\label{vr}
Radial velocities available for the new clusters. } 
\begin{flushleft}
\begin{tabular}{lrrrr}
\hline\hline
\noalign{\smallskip}
\hbox{Cluster} & \hbox{RV (km.s$^{-1}$)} & \hbox{Reference} \\ 
\noalign{\smallskip}
\hline
\noalign{\smallskip}
VVV-CL001 & $-324.9\pm0.08$ & 1 \\
FSR1758 & $226.8\pm0.08$ & 2 \\
VVV-CL002 & $-27.3\pm0.10$ & 3 \\
Patchick 99 & $92\pm10$ & 4 \\
Gran 1 &  $76.98\pm3.62$ & 5 \\
Gran 2 &  $61.24\pm2.70$  & 5  \\
Gran 3 &  $91.57\pm5.97$   & 5 \\
Gran 4 &  $-265.28\pm 3.92$  & 5  \\
Gran 5 &  $-59.19\pm 4.93$  & 5    \\
\hline
\noalign{\hrule\vskip 0.1cm} 
\hline                  
\end{tabular}
\end{flushleft}
References: 1 \citet{fernandeztrincado21a}; 
2 \citet{villanova19}; 3  \citet{minniti24}; 4  \citet{butler24}; 5 \citet{gran24}.
\end{table}   

\citet{minniti21a} report another 13 cluster candidates that should be further studied:
VVV-CL154, and Minniti 20, 39, 40, 46, 47, 54, 55, 56, 57, 58, 59, and 60,
detected in the VVV survey.

\begin{table*}
\caption{Bulge GCs from \citet{bica16}, with distances, ages, and metallicity values from the literature, in order of right ascension. UKS~1 has been added (see text).
}
\label{bica16}
\scalefont{0.7}
\begin{flushleft}
\begin{tabular}{lccccccc@{}c@{}c@{}cc@{}c@{}c@{}c@{}c@{}c@{}c@{}}
\hline
\hline
\noalign{\smallskip}
\noalign{\smallskip}
   Cluster & $\ell$ & $b$ & \multicolumn{5}{c}{Distance to the Sun (kpc)} & M$_{\rm V}$ & [Fe/H] & Ref. & Age & Ref. \\
           & ($^\circ$) & ($^\circ$) & Adopted (PV20) & H10 & BV21 & Our group  & Ref.  & (mag.) & (dex) &  & (Gyr) &  \\ 
\noalign{\smallskip}
\hline
\hline
\noalign{\smallskip}
Terzan 3$^{I}$         &345.08 & 9.19& $8.20\pm0.82$  & 8.20  & 7.644$^{+0.318}_{-0.304}$  & 6.50$\pm 1.0$ & 1 & -4.82  
&-1.01$\pm 0.08$ & 1, 2 & -- & -- \\
ESO452-SC11$^{B,I}$ &351.91 & 12.10 &$6.50\pm0.65$   &  8.30  & 7.389$^{+0.202}_{-0.196}$  & -- & -- & -4.02   & 
-0.81$\pm 0.13$ & 3 & -- & -- \\
NGC 6256$^{B,I}$  & 347.79 & 3.31& $6.40\pm0.64 $ & 10.30 & 7.252$^{+0.299}_{-0.288}$  &  6.4$\pm 1.0$ & 2  & -7.15  & 
-1.05$\pm 0.13$ &  1 & $12.9\pm1.0$ & 1 \\
NGC 6266$^{B,I}$  & 353.58& 7.32& $6.41\pm0.12$  & 6.80  & 6.412$^{+0.105}_{-0.104}$  & 6.6$\pm 0.5$ & 3   & -9.18  &
-1.18$\pm 0.07$ & 4 & $11.6\pm0.6$ & 2 \\ 
NGC 6304$^{B,I}$  & 355.83 & 5.38 & $6.28\pm0.11$  & 5.9   & 6.152$^{+0.150}_{-0.146}$  & 6.07$\pm 0.009$ & 4 & -7.30  &
-0.37$\pm 0.07$ & 4  & $12.3\pm0.8$ & 3 \\ 
NGC 6316$^{I}$        & 357.18&5.76& $11.60\pm1.16$ & 10.40 & 11.152$^{+0.393}_{-0.382}$ & -- & -- & -8.34  &
-0.50$\pm 0.06$ & 1 & -- & -- \\ 
NGC 6325$^{I}$         & 0.97&8.00 & $7.80\pm0.78$  & 7.80  & 7.533$^{+0.330}_{-0.314}$  & 6.9$\pm 1.0$ & 5 & -6.96  & 
-1.38$\pm 0.10$ & 1, 2 & $12.5\pm0.9$ & 1 \\
NGC 6342$^{B,I}$  &4.90 &9.73 & $8.43\pm0.84$  & 8.50  & 8.013$^{+0.233}_{-0.226}$  & -- & --  & -6.42  &
-0.55$\pm 0.05$ & 5 & $11.5\pm1.3$ & 1 \\
NGC 6355$^{I}$         & 359.58& 5.43& $8.70\pm0.87$  & 9.20  & 8.655$^{+0.224}_{-0.220}$  & 8.54$\pm 0.19$ & 6 & -8.07  & 
-1.39$\pm 0.08$ & 6  & $13.2\pm1.0$ & 1,4 \\
Terzan 2$^{I}$         &356.32 & 2.30& $7.50\pm0.75$  & 7.50  & 7.753$^{+0.332}_{-0.318}$  & 7.7$\pm 0.6$ & 7 & -5.88  &
-0.85$\pm 0.04$ & 7  & -- & -- \\
Terzan 4$^{I}$         &356.02 & 1.31& $6.70\pm0.67$  & 7.20  & 7.591$^{+0.315}_{-0.302}$  & 8.0$\pm 0.7$ & 8  & -5.74  &
-1.40$\pm 0.05$ & 7,8  & -- & -- \\
HP 1   $^{B,I}$    &357.42 &2.12 &$ 6.59\pm0.16$  & 8.20  & 6.995$^{+0.144}_{-0.141}$  & 6.59$^{+0.15}_{-0.15}$ & 9 & -6.46  &-1.06,-1.20$\pm 0.10$ & 9,7 & $12.8\pm0.9$ & 5 \\
Liller 1 $^{B,I}$  &354.84 & -0.16& $8.20\pm0.82$  & 8.20  & 8.061$^{+0.353}_{-0.338}$  & 6.5$\pm 1.0$ & 8 &  -7.32 &
-0.3$\pm 0.22$ & 10** & $12.0\pm1.5$ & 6 \\
Terzan 1$^{I}$         & 357.57&1.00 & $6.70\pm0.67$  & 6.70  & 5.673 $^{+0.175}_{-0.170}$  & 5.2$\pm 0.5$ & 10 & -4.41  &
-1.06$\pm 0.13$ & 1 & -- & -- \\
Ton 2$^{I}$            &350.80 &-3.42 & $6.40\pm0.64$  & 8.20  & 6.987$^{+0.344}_{-0.326}$  & 6.4$\pm 1.0$ & 11 & -6.17  &
-0.73$\pm 0.13$ & 1  & -- & -- \\ 
NGC 6401$^{B,I}$  & 3.45 & 3.98 & $7.70\pm0.77$   & 10.6  & 8.064$^{+0.238}_{-0.234}$   & 12.0$\pm 1.0$ & 12 &  -7.90 &
-1.12$\pm 0.07$ & 1  & $13.2\pm1.2$ & 1 \\
Palomar 6$^{I}$       &2.09 & 1.78& $5.80\pm0.58$  & 5.80  & 7.047$^{+0.463}_{-0.433}$  & 7.67$\pm 0.19$ & 13 & -6.79  & 
-1.10$\pm 0.09$ & 11  & $12.4\pm0.9$ & 7 \\
Djorg 1$^{I}$         &356.67 &-2.48 & $9.30\pm0.50$  & 13.70 & 9.879$^{+0.671}_{-0.628}$ & 9.3$\pm 0.5$ & 14 & -6.98  &
-1.54$\pm 0.13$ & 1  & -- & -- \\ 
Terzan 5$^{I}$        &3.81 & 1.67& $5.50\pm0.51$  & 6.90  & 6.617$^{+0.150}_{-0.148}$  & 4.6$\pm 0.9$ & 8 & -7.42  & 
-0.3$\pm 0.12$ & 8,12** & $12.0\pm1.0$ & 8 \\
NGC 6440$^{B,I}$   &7.73 & 3.80& $8.24\pm0.82$  & 8.5   & 8.248$^{+0.248}_{-0.241}$  & 8.47$\pm 0.4$ & 15 & -8.75  & 
-0.24$\pm 0.05$  & 1, 2 & $13.0\pm1.5$ & 9 \\
Terzan 6$^{I}$         &358.57 &-2.16 & $6.80\pm0.68$  & 6.80  & 7.271$^{+0.360}_{-0.343}$  & 7.0$\pm 1.0$ & 16  & -7.59  & $-0.43\pm 0.08$ & 13 & -- & --  \\ 
UKS 1$^{I}$        &5.13 &-0.76 & $11.1\pm1.8$   & 7.8   & 15.581$^{+0.570}_{-0.550}$   & 10.6$\pm 1.5$ & 8 & -6.91  &
-0.98$\pm 0.11$ & 14 & $13.0\pm1.2$ & 10 \\ 
Terzan 9$^{I}$        &3.60 &-1.99 & $7.10\pm0.71$  & 7.10  & 5.770$^{+0.356}_{-0.333}$  & 4.9$\pm 1.0$ & 17 & -3.71  &-
1.10,-1.40$\pm 0.15$ & 15,7 & -- & -- \\ 
ESO456-SC38$^{B,I}$ 
&2.76 &-2.50 & $8.75\pm0.20$  & 6.30  & 8.764$^{+0.178}_{-0.174}$  & 8.75$\pm 0.12$ & 18 & -7.00  &
--1.07$\pm 0.09$   & 7 & $12.7\pm0.7$ & 11 \\
Terzan 10$^{I}$        &4.42 &-1.86 & $10.3\pm1.0$   & 5.8   & 10.212$^{+0.412}_{-0.396}$ & 10.3$\pm 1.0$ & 14   & -6.35  &
-1.47$\pm 0.02$ & 13 & -- & -- \\ 
NGC 6522$^{B,I}$  &1.02 &-3.93 &$ 7.40 \pm 0.19$& 7.70  & 7.295$^{+0.211}_{-0.205}$ & 7.62$\pm 0.17$ & 19 &  -7.65 & 
-1.05$\pm 0.11$ & 16 & $12.8\pm1.0$ & 12 \\
NGC 6528$^{B,I}$  &1.14 &-4.17 &$ 7.70 \pm 0.77$& 7.90  & 7.829$^{+0.239}_{-0.232}$  & 7.7$\pm 1.0$ & 8 & -6.57  & 
-0.17$\pm 0.07$ & 17,9 & $12.7\pm1.0$ & 13 \\ 
NGC 6539$^{I}$       &20.80 &6.78 &$ 7.85 \pm 0.66$& 7.8   & 8.165$^{+0.395}_{-0.379}$  & --  & --  & -8.29  &
-0.55$\pm 0.06$ & 1 & -- & -- \\
NGC 6540$^{B,I}$  &3.29 &-3.31 & $5.20\pm0.52$  & 5.30  & 5.909$^{+0.279}_{-0.265}$  & 3.50$\pm 0.4$ & 20 & -6.35  & 
-1.06$\pm 0.06$  & 7 & -- & -- \\
NGC 6553$^{B,I}$  & 5.25&-3.02 & $6.75\pm0.22$  & 6.0   & 5.332$^{+0.128}_{-0.125}$  & 5.2$\pm 1.0$ & 21 & -7.77  &
-0.27$\pm 0.09$ & 18 & $12.0\pm2.0$ & 14 \\ 
NGC 6558$^{B,I}$  &0.20&-6.03 & $8.26\pm0.53$  & 7.40  & 7.474$^{+0.294}_{-0.282}$  & 6.3$\pm 1.0$ & 22 & -6.44  &
-1.16$\pm 0.08$  &19 & $12.3\pm1.1$ & 1 \\
BH 261$^{B,I}$    & 3.36 & -5.27 &      $6.50\pm 0.65$& 6.50  & 6.115$^{+0.265}_{-0.253}$  &  -- & --  & -4.06  &
-1.09$\pm 0.05$ & 13 & -- & -- \\ 
Glimpse02        &14.13 &-0.64 &     --         & 5.50  &  --    & --   & -- & --   &
-1.08$\pm 0.13$ & 20 & -- & -- \\ 
Mercer~5         &17.59&-0.86 &  $5.50\pm 0.55$&  --   & 5.466$^{+0.523}_{-0.476}$ &  --  & -- &  -- & 
-0.86$\pm 0.12$ & 20 & -- & -- \\
NGC 6624$^{B,I}$  &2.79 &-7.91 & $8.43\pm 0.11$ & 7.9   & 8.019$^{+0.108}_{-0.107}$  &  8.19$\pm 0.15$ & 4 & -7.49  &
-0.36$\pm 0.12$ & 21 & $11.7-12.0\pm0.6$ & 3,15 \\
NGC 6626$^{B,I}$  &7.80 &-5.58 & $5.34\pm0.17$  & 5.5   & 5.368$^{+0.099}_{-0.098}$  & 5.6$\pm 0.5$ & 3 & -8.16  &
-1.29$\pm 0.01$  & 22 & $12.8\pm1.0$ & 12 \\
NGC 6637$^{B,I}$  &1.72&-10.27& $8.80\pm0.88$  & 8.80  & 8.900$^{+0.106}_{-0.104}$  & 8.75$\pm 0.12$ & 4  & -7.64  &
-0.59$\pm 0.07$ & 4 & $12.3\pm0.6$ & 3 \\ 
NGC 6638$^{I}$        & 7.90 & -7.15 & $10.32\pm1.03$ & 9.40  & 9.775$^{+0.347}_{-0.334}$ & 9.6$\pm 0.5$ & 3   & -7.12  &
-0.99$\pm 0.07$ & 4 & 12.0 & 2 \\ 
NGC 6642$^{B,I}$  &9.81 &-6.44 & $8.10\pm0.81$  & 8.10  & 8.049$^{+0.204}_{-0.198}$  & 8.2$\pm 0.7$ & 3  & -6.66  &
-1.11$\pm 0.14$ & 4 & $12.7\pm1.1$ & 1 \\
NGC 6652$^{I}$         &1.53 &-11.38 & $9.51 \pm0.12$ & 10    & 9.464$^{+0.139}_{-0.137}$ & 9.34$\pm 0.18$ & 4  & -6.66  &
-0.76$\pm 0.14$ & 4 & $12.7\pm0.7$ & 3 \\ 
NGC 6717$^{B,I}$  &12.88 &-10.90 & $7.14\pm0.10$  & 7.10  & 7.524$^{+0.133}_{-0.131}$  & 7.3$\pm 0.5$ & 3 & -5.66  &
-1.26$\pm 0.07$ & 4 & $13.5\pm0.8$ & 3 \\ 
NGC 6723$^{I}$         & 0.07&-17.30 & $8.17\pm0.11$  & 8.70  & 8.267$^{+0.100}_{-0.099}$  & 8.09$\pm 0.11$ & 4  & -7.83  &
-1.01$\pm 0.05$ & 24 & $12.6\pm0.6$ & 3 \\ 
\noalign{\smallskip}
\hline
\noalign{\smallskip}
\hline
\end{tabular}
\end{flushleft}
Symbols:
$^{B}$ identified as a bulge cluster by \citetalias{perez20}, $^{I}$ identified as in situ by \citet{belokurov24}.
References to distance by our group: 
1 \citet{barbuy98}; 
 2 \citet{ortolani99a}; 
 3 \citet{oliveira22}; 4 \citet{oliveira20}; 5 \citet{ortolani00}; 
 6 \citet{souza23};  
 7 \citet{ortolani97}; 
 8 \citet{ortolani07}; 
 9 \citet{kerber19}; 
 10 \citet{ortolani99b}; 
 11 \citet{bica96}; 
 12 \citet{barbuy99a};
 13 \citet{souza21};
 14 \citet{ortolani19a}; 
 15 \citet{ortolani94}; 
 16 \citet{barbuy97}; 
 17 \citet{ortolani99c}; 
 18 \citet{ortolani19b}; 
 19 \citet{kerber18};
 20 \citet{bica94};
 21 \citet{guarnieri98}; 
 22 \citet{rich98}.
References to metallicity:  1 \citet{vasquez18} from CaT, calibrated to \citet{dias16};
2 \citet{saviane12}; 3 \citet{simpson17}; 4 \citet{carretta09}; 5 \citet{johnson16}; 6 \citet{souza23};
7 \citet{geisler21};
8 \citet{origlia04};
9 \citet{barbuy16}; 10 \citet{crociati23} from MUSE; 11 \citet{souza21};  12 \citet{massari14}; 
13 \citet{geisler23} from CaT; 
14 \citet{fernandez20};
15 \citet{ernandes19} from MUSE;
16 \citet{barbuy21a} and \citet{fernandez19};
17 \citet{carretta01}, \citet{zoccali04}, and \citet{munoz18};
18 \citet{barbuy99b}, \citet{melendez03}, \citet{alvesbrito06}, \citet{origlia02}, and \citet{montecinos21};
19 \citet{barbuy18b} and \citet{gonzalez23};
20 \citet{penaloza15}; 21 \citet{husser20} from MUSE; 22 \citet{villanova17}; 23 \citet{crestani19};
24 \citet{fernandeztrincado21b}. ** For Terzan 5 and Liller 1, we adopt a mean weighted to the most important multi-population component.
References of age: 1 \citet{cohen21} with relative ages; 2 \citet{meissner06}; 3 \citet{oliveira20}; 4 \citet{souza23}; 5 \citet{kerber19}; 6 \citet{ferraro21}; 7 \citet{souza21}; 8 \citet{ferraro16}; 9 \citet{pallanca21}; 10 \citet{fernandez20}; 11 \citet{ortolani19b}; 12 \citet{kerber18} with a mean age between Dartmouth and BasTI models; 13 \citet{renzini18}; 14 \citet{bruzual97}; 15 \citet{saracino16}.
\end{table*}


\section{Previously identified bulge sample}
\label{sec:3}

As was mentioned above, \citet{bica16} presented a review of GCs that have characteristics of belonging to the
Galactic bulge, specified as: a) an angular distance to the Galactic center lower than 20$^{\circ}$, 
b) a distance to the Galactic center shorter than $3.5$\,kpc, and c) a metallicity of [Fe/H]$>-$1.5\,dex.
Table~\ref{bica16} revises the data available for the 42 GCs identified as
belonging to the bulge populations by \citet{bica16}. In particular, metallicities have been derived for many of them in recent years using
high-resolution spectroscopy. 
We note that for Terzan 5 we adopt the mean of $\rm{[Fe/H]}=-0.3$\,dex, corresponding to the dominant population
of the various populations, which are estimated to have $\rm{[Fe/H]} = 0.25,-0.3,-0.8$\,dex \citep{massari14};
this is compatible with \citet{origlia04}. Likewise,
for Liller 1 we also adopt a mean of $\rm{[Fe/H]}=-0.3$\,dex, which is that of the dominant population and a mean from
$-0.48$ and $+0.27$\,dex \citep{crociati23}, and compatible with \citet{origlia02}.
The coordinates of these clusters, not reported in Table~\ref{bica16}, can be found in \citetalias{harris10}; likewise the quantities
A$_{K_s}$, $\mu_{\alpha}^{*}$, and $\mu_{\delta}$ are given by \citet{vasiliev21}.

Distances are reported and given in four different sources, which is useful because there are some discrepant values in the literature.
The final distances are adopted from \citetalias{perez20}.

The ages were also retrieved from the literature, and are provided in Table~\ref{bica16}.
Many of the sample objects were observed after the 1990s
and different techniques were used to derive their ages.
For the sake of homogeneity, we only considered the studies with resolved photometry.
Most of the CMDs are from \textit{HST} observations. Exceptions are
the deep near-infrared data in $K_{S}$ versus $J-K_{S}$
from \citet{kerber19} and \citet{saracino16} that used the Gemini
imager with adaptive optics, and the observations of UKS~1 with the VVV
survey; the latter is not deep, reaching just above the turnoff.
The isochrone fitting procedures in \citet{kerber19}, \cite{ortolani19b}, \cite{oliveira20}, \cite{fernandez20}, and \cite{souza21, souza23}
employed a statistical Bayesian method with the \texttt{SIRIUS} code \citep{souza20}, based on grids of Dartmouth \citep{dotter08} and/or BaSTI isochrones \citep{pietrinferni06}, and
a similar procedure was applied by \citet{kerber18}. 
\citet{cohen21} used the fiducial lines of both the studied clusters and well-known clusters with reliable age determinations to derive their relative ages.
\citet{bruzual97} employed the early Padova isochrones \citep{bressan93}, whereas 
\citet{renzini18}  employed the Victoria-Regina code and isochrones 
\citep{vandenberg14}.
\citet{saracino16}, \citet{ferraro21}, and \citet{pallanca21}
employed fiducial lines and $\chi^2$ minimization to fit isochrones, employing the models
mentioned above.


Age determinations using other methods, such as the magnitude difference between the horizontal branch and the main sequence turnoff,
integrated spectra models, and $M_V$ calibrations of RR Lyrae stars, were not considered.
The ages range from $11.5$ to $13.5$\,Gyr, with an average age of 12.5\,Gyr, in agreement with a mean of 12.3\,Gyr from \citet{oliveira20}.
Fifteen clusters with no age determination are mostly low-mass and very reddened.

\begin{SCfigure*}[0.5]
\centering
\includegraphics[width=1.5\columnwidth]{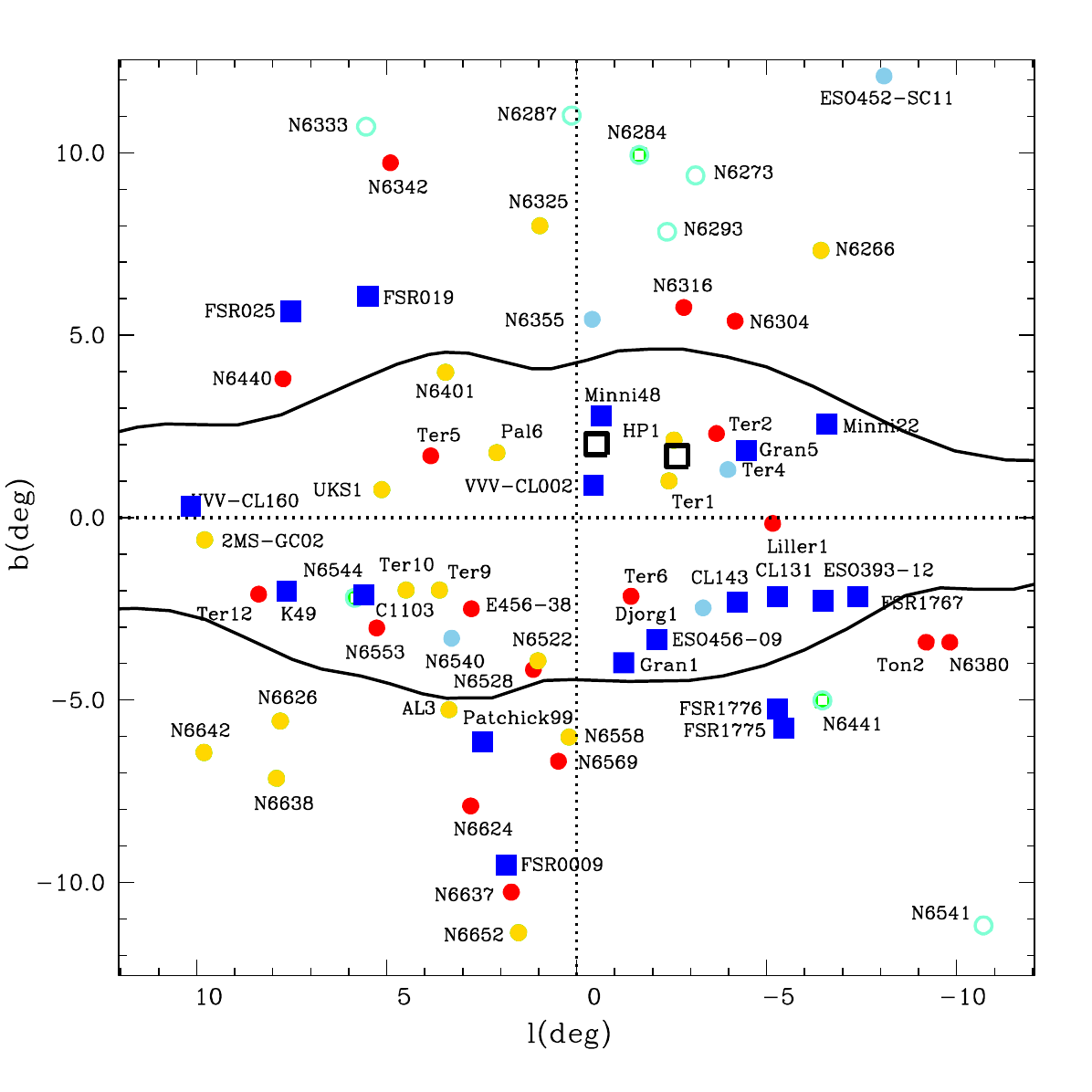}
\caption{Location in  Galactic coordinates of bulge clusters, including a) projected bulge globular clusters cf. \citet{bica16}, represented by filled circles,
with [Fe/H]$>$-0.8 (red); -1.3$<$[Fe/H]$<$-0.8 (gold); [Fe/H]$<$-1.3 (blue);  halo clusters:
aquamarine open circles; cluster farther from the Galactic center than R$_{\rm C}$$>$4.5 kpc: green open squares;
b) candidate clusters (open black squares); c)  newly confirmed clusters located in the region (blue filled squares).
The Galactic
center is illustrated by the blue filled circle. Contours correspond to COBE/DIRBE outline of the
bulge from \citet{weiland94}, and adapted from \citet{jonsson17}.}
\label{census}
\end{SCfigure*}

In Table \ref{bica16} we identify the clusters classified as bulge members according to their orbits by
\citet{perez20}, identified by a $^{B}$. We also indicate, in Tables \ref{census} and \ref{bica16}, the classification of having been formed in situ or accreted,
based on the total energy, $E$, and $z$ component of the orbital angular momentum
and further calibrated using the [Al/Fe] abundance ratio, by \citet{belokurov24}, indicated by $^{I}$ or $^{A}$,
corresponding to ``in situ'' or ``accreted.''

In Figure~\ref{census} the location of the bulge clusters is plotted in galactic coordinates, $\ell$ and $b$, in degrees.
This figure is an update of figure 3 from \citet{barbuy18a}, including 
a) the bulge GCs, \citet{bica16} 
b) candidate clusters, and c) newly confirmed clusters.
All of the objects from lists b) and c) are new relative to a).
UKS~1 is now included as a previously known cluster: it was not included as such in \citet{bica16} for the reason that its 
distance was a limiting case. It still is, but it is a typical bulge cluster in terms of CMD, which justifies its inclusion
in the list; its true distance remains an open question.


There are two cases of clusters that had previously been identified at the time of the \citet{bica16} review but that had essentially no data, 
and that have since been observed and characterized in recent surveys, as is shown in Table~\ref{logbook}.
These clusters are VVV-CL002 and Kronberger 49, and they are plotted in the figures as new clusters.
Therefore, we consider hereafter 42 well-known clusters, comprising those from \citet{bica16},
except for VVV-CL002 and Kronberger 49, which we took out from Table~\ref{bica16}, in addition to UKS~1.

\section{Parameters of the new sample}
\label{sec:4}

\subsection{Metallicity and mass}

\begin{figure*}
\centering
\includegraphics[width=0.48\textwidth]{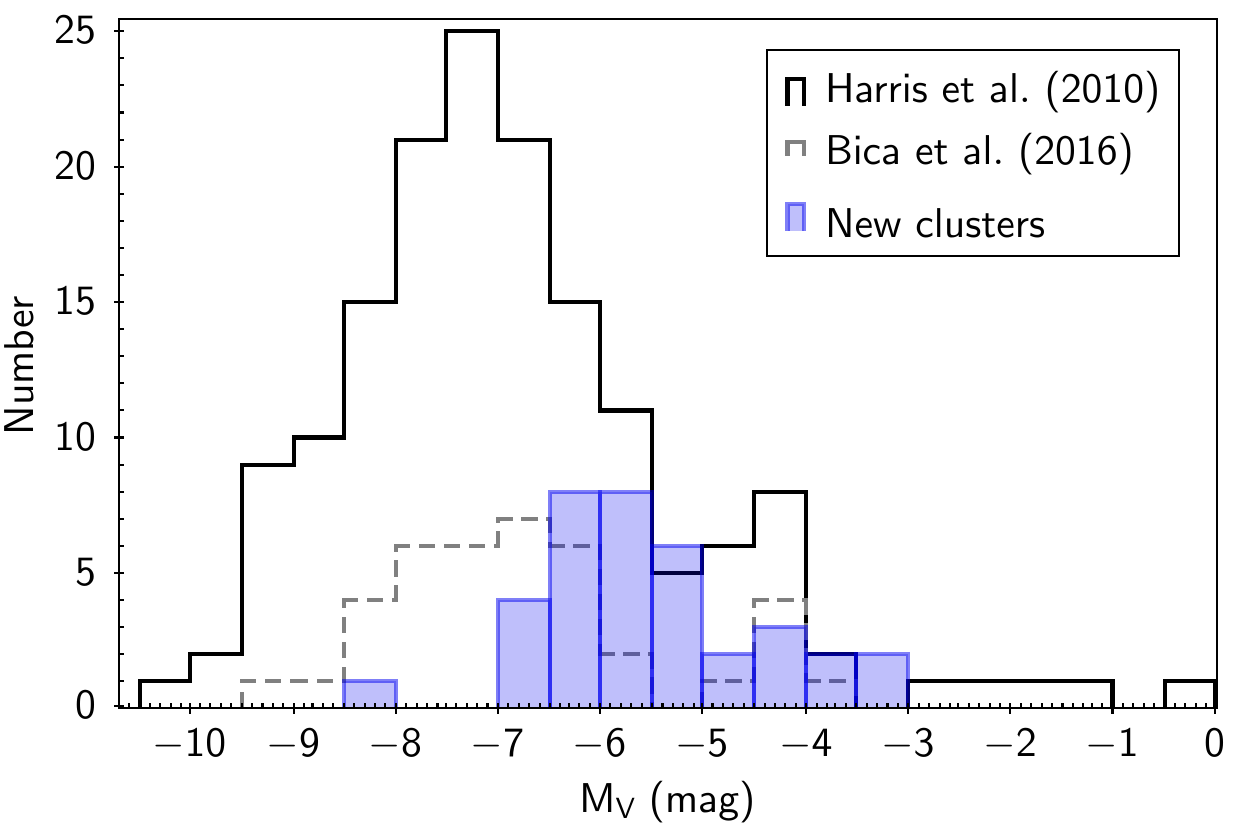}
\includegraphics[width=0.48\textwidth]{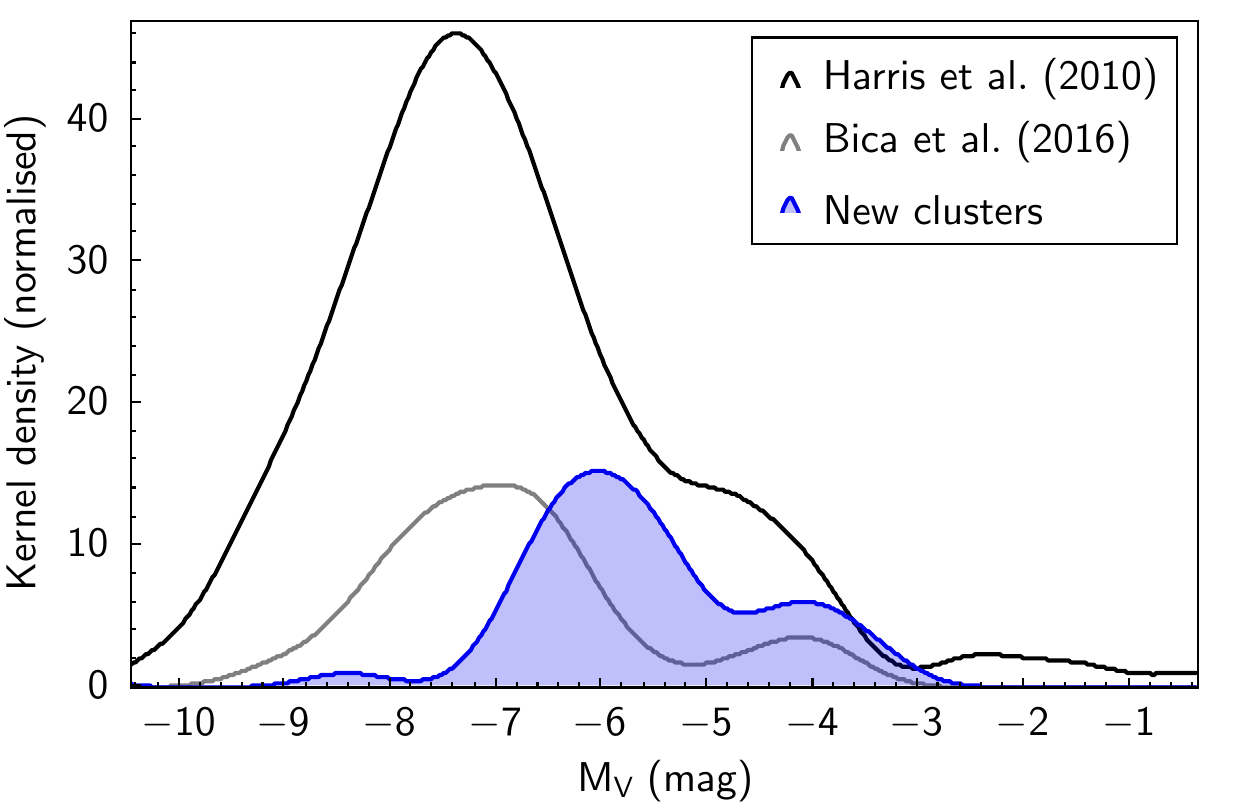}
\caption{
Absolute magnitude, $M_V$, of the new cluster sample compared to the literature.
\textit{(Left panel:)} Histogram of the 36 of the 41 new clusters from Table~\ref{logbook} with available $M_V$ (solid blue histogram), compared to \citetalias{harris10} (156 clusters, solid black histogram) and bulge clusters  \citet[42 clusters]{bica16} plus UKS~1 (dashed gray histogram).
\textit{(Right panel:)} Kernel density estimation of the three distributions, with a Gaussian kernel and normalized $y$ axis. A low-luminosity tail is seen in the three samples between $M_V\sim -5$ and $-4$\,mag.}
\label{absMag} 
\end{figure*}

The sample clusters have lower masses than the well-known GCs, as
is indicated by their integrated absolute
magnitudes, with most of them with $M_{\rm V}$ fainter than $-7.0$\,mag. This is clearly seen in
Figure \ref{absMag}, where only clusters that have an $M_{\rm V}$ value
available  are included.
This figure shows that the new clusters all have low masses relative to the other two samples.
It is therefore unlikely that massive clusters are still missing, but there could still be
massive clusters with low density that remain undetected. Other low mass clusters probably also remain undetected due to crowding and high extinction.
It can be seen in the right panel of Figure \ref{absMag} that the low-luminosity tail of the globular cluster luminosity function (GCLF) is non-Gaussian, given an extended tail of low-mass clusters 
\citep[see also the discussion on this matter by][]{jordan07}.

\begin{figure*}
\centering
\includegraphics[width=0.48\textwidth]{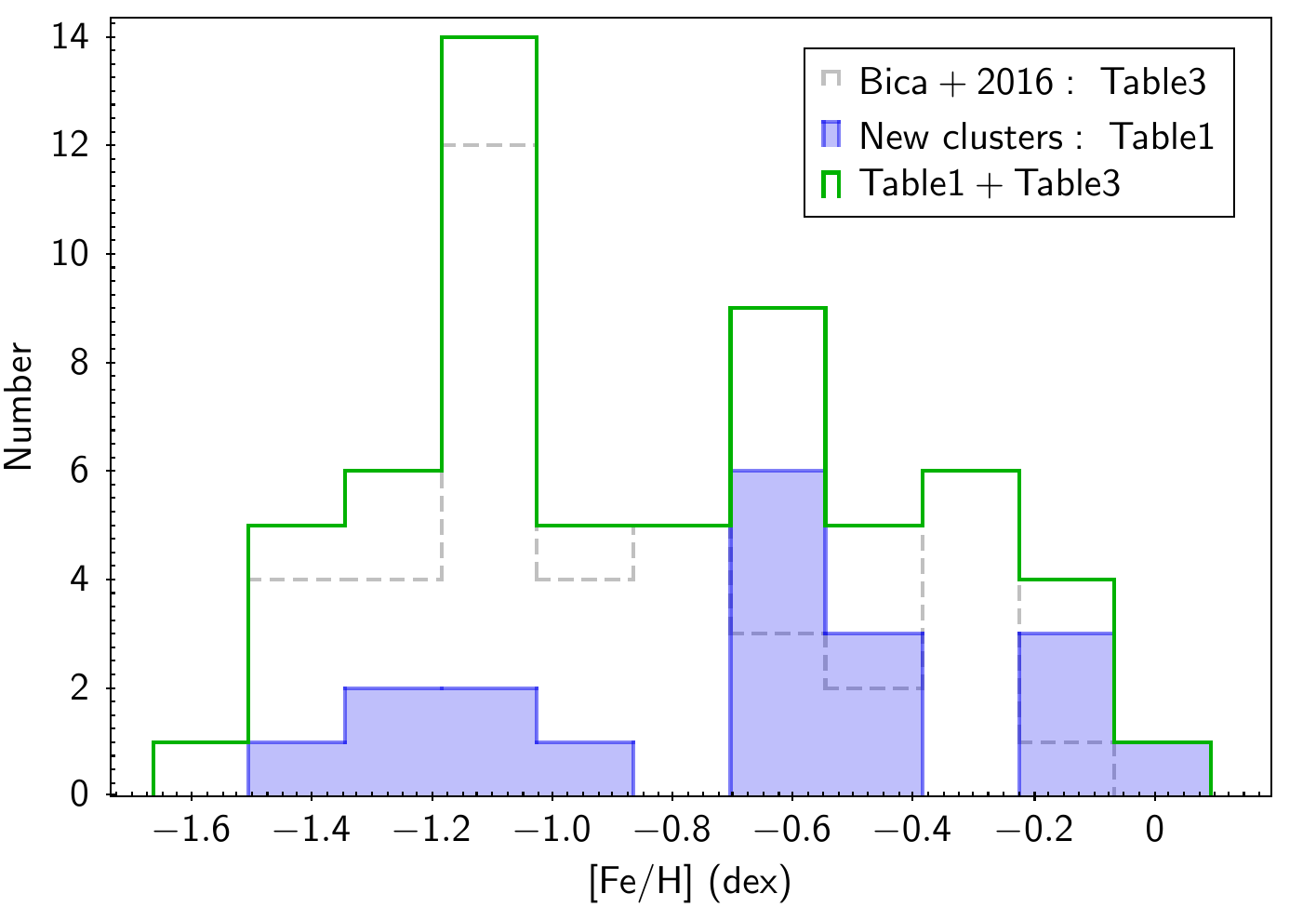}
\includegraphics[width=0.48\textwidth]{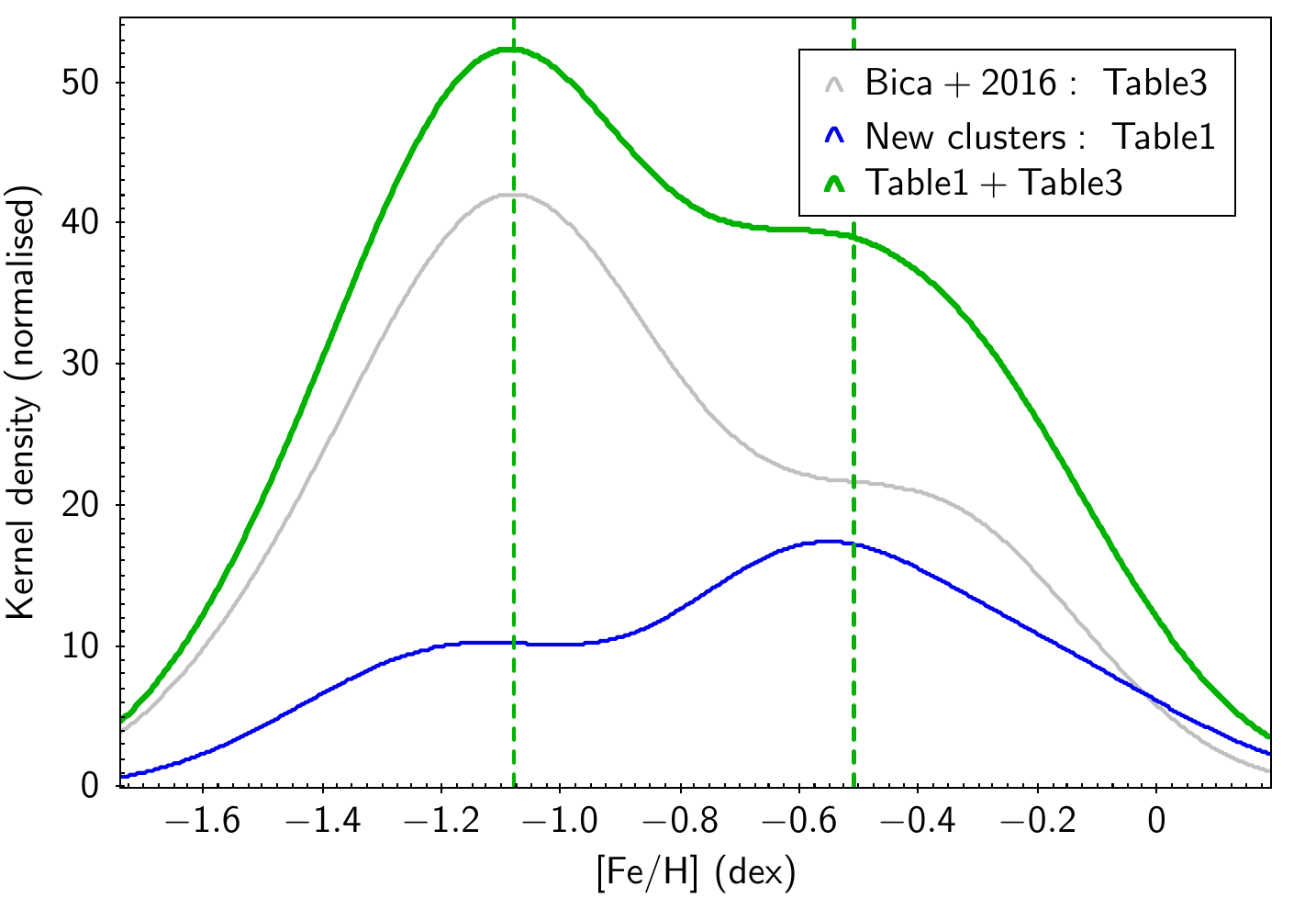}
\caption{Metallicities of full sample of bulge clusters.
\textit{(Left panel:)} Histogram of metallicity including the 42 well-known bulge clusters from \citet{bica16}
and the new sample clusters, excluding clusters classified as belonging to the halo and disk,
those that have $\rm{[Fe/H]}<-1.5$\,dex,  
and clusters that have distances out of the range of $4 < \rm{d}_{\odot} < 11$\,kpc.
A histogram combining both samples is also shown in green. \textit{(Right panel:)} Kernel density estimation, showing the distribution peaks independently of the bin size: \citet{bica16} with peaks in $-1.05$ and $-0.40$\,dex, the new cluster sample with peaks in $-1.20$ and $-0.55$\,dex, and the combined sample with a higher peak consistent with \citet{bica16} but with a smaller peak shifted to a more metal-poor value (dashed lines: $-1.08\pm0.35$ and $-0.51\pm0.25$\,dex).}
\label{metallicity} 
\end{figure*}

Figure \ref{metallicity} shows the metallicity distribution of bulge clusters, including the well-known bulge clusters reported in Table \ref{bica16}, together with
the new sample clusters listed in Table \ref{census},
excluding the halo, halo intruder, and disk clusters,
clusters that have
$\rm{[Fe/H]}\simless-1.5$\,dex (in order to be compatible with the selection criteria of
bulge clusters defined in \citealt{bica16}), 
and clusters that have distances out of the range of $4\simless\rm{d}_{\odot}\simless 11$\,kpc.

The left panel shows the histogram of metallicity and the right panel kernel density estimations.
From the right panel of Fig. \ref{metallicity} the combined bulge sample (Tables \ref{census} and \ref{bica16})
show peaks at $\rm{[Fe/H]} \sim -1.08\pm0.35$ and $-0.51\pm0.25$\,dex.
The more metal-poor peak is in agreement with that of \citet{bica16}, whereas the more metal-rich peak is shifted by $\sim 0.1$\,dex compared to \citet{bica16} due to the large number of new clusters with $\rm{[Fe/H]}\sim-0.6$\,dex.
The new clusters alone have peaks at $\rm{[Fe/H]} \sim  -0.55$\,dex, coincident with the full sample,
and at a lower metallicity of $\rm{[Fe/H]} \sim -1.2$\,dex, but with low statistics.

The seven new halo clusters are all more metal-poor than $\rm{[Fe/H]} < -1.3$\,dex,
except for VVV-CL003. VVV-CL001 is more metal-poor than any halo cluster,
with $\rm{[Fe/H]} \sim -2.45$\,dex \citep{fernandeztrincado21a}, and it has been suggested, owing to its orbit, that it belongs to a massive dwarf galaxy such as Gaia-Enceladus-Sausage \citep[GES;][]{belokurov18, helmi18} or
Sequoia \citep[Seq.;][]{massari19}
that merged with the Galaxy at an early time.
Another cluster with such low metallicity in the bulge is ESO 280-SC06
\citep{ortolani00}, which was recently studied spectroscopically, revealing a metallicity
 of $\rm{[Fe/H]}=-2.48$\,dex by \citet{simpson18}, and reported to be associated with the
Gaia-Enceladus-Sausage \citep{massari19} dwarf galaxy.

The bulge field surveys GIBS \citep{zoccali17} and ARGOS \citep{ness13}
show peaks at $\rm{[Fe/H]}\sim+0.3$, $-0.4$\,dex and $\rm{[Fe/H]}\sim+0.15$, $-0.25$, and $-0.7$\,dex, respectively.
The combined sample of bulge clusters
shows  compatibility between bulge clusters and the field at $\rm{[Fe/H]}\sim-0.5$\,dex from \citep{zoccali17},
and $\rm{[Fe/H]}\sim-0.7$\,dex from \citep{ness13}.
The more metal-poor peak at $\rm{[Fe/H]}\sim-1.1$\,dex already pointed out in \citet{bica16} is confirmed here.


\subsection{Ages}

Regarding ages, only 28 of the new sample clusters have values of age and
metallicity reported in the literature. 

The ages were obtained from the visual fitting of isochrones. The CMDs are not deep,
with those from VVV data reaching $K_{S}$ $\sim 17-18$\,mag. in the best cases;
those from WISE data are shallow, with $K_{S} \sim 14-15$\,mag.
\citet{garro22a} give an age uncertainty of $\pm 2$\,Gyr, while \citet{fernandeztrincado22} give $\pm 3$\,Gyr. We can adopt a general
uncertainty of $\pm 2 - 3$\,Gyr.
In order to have a comparison sample of clusters formed in situ, and another sample formed 
ex situ, we adopt the selection by \citet{forbes10} and \citet{forbes20}, 
with data updated in \citet{kruijssen19}, and \citet{limberg22}.
These samples were intended to verify an AMR of the new clusters from 
Table \ref{census}.

\begin{figure}
\centering
\includegraphics[width=8cm]{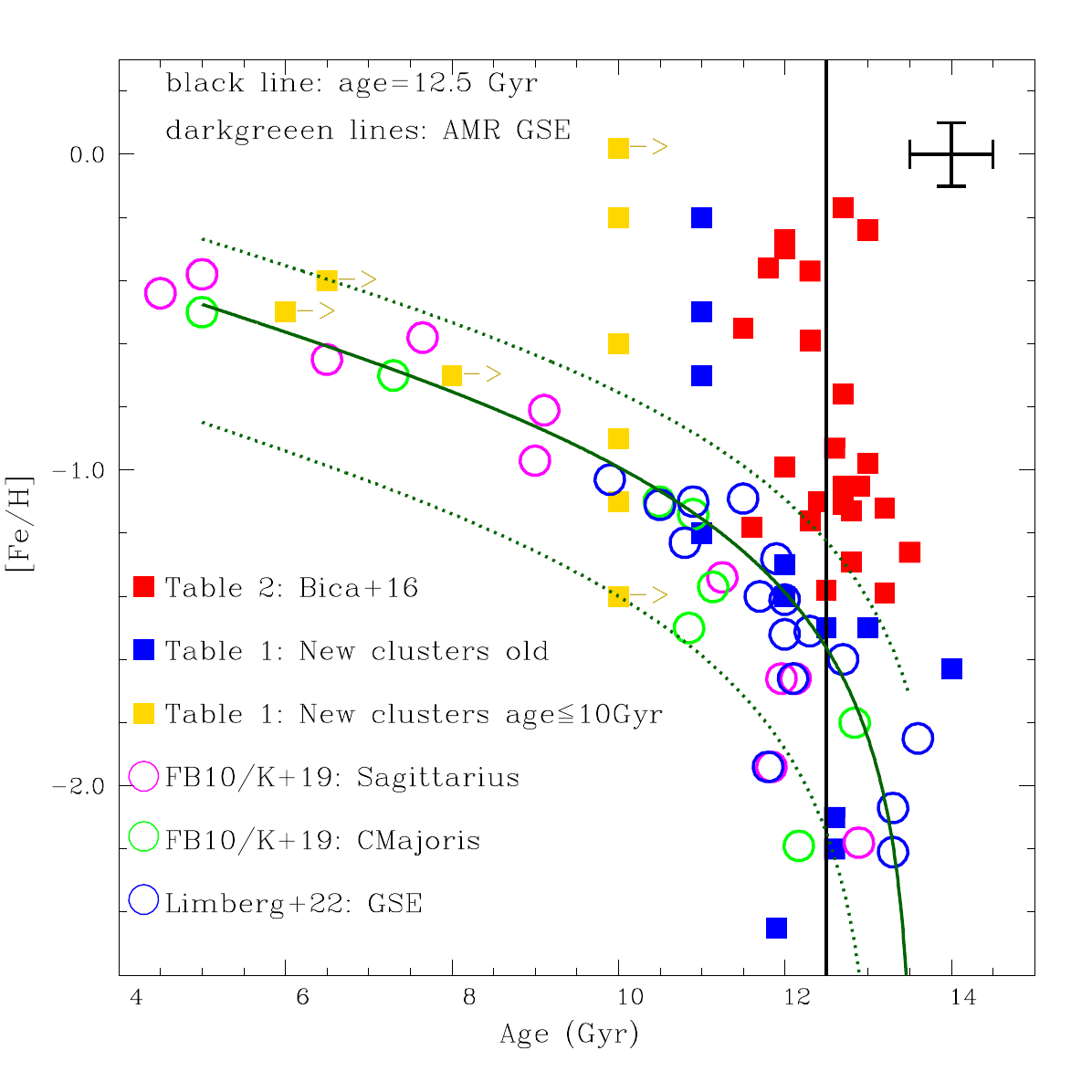}
\caption{Metallicity vs. age for the new clusters that have these values available,
compared with the \cite{bica16} sample,
and with the Sagittarius and Canis Majoris clusters identified in \citet{forbes10}
labeled as FB10, with updates from \citet{kruijssen19} labeled as K+19, 
and Gaia-Enceladus-Sausage with clusters identified in \citet{limberg22}.
Bulge clusters from Table~\ref{bica16} are represented by filled red squares; old clusters from Table~\ref{logbook} by filled blue squares; clusters from Table~\ref{logbook} younger than 10 Gyr by filled gold squares; clusters from Sagittarius by open magenta circles; clusters from Canis Majoris by open green circles; clusters from GSE by open blue circles; ages of $12.5\pm0.3$\,Gyr by black lines; and the model
and uncertainties from \citet{forbes10} by green lines.
The old bulge clusters are all old and show no AMR,
as is indicated by the straight black line with an age of 12.0\,Gyr.
Dark green curves represent the best fit of the AMR to the GSE sample from \citet{limberg22},
and $\pm$3$\sigma$ curves.
The lower age limits for five of the new clusters are represented by arrows. The error bar in the upper right corner indicates the typical errors of
0.2 dex in [Fe/H] and 1.0 Gyr in age.}
\label{age-met} 
\end{figure}

Figure \ref{age-met} compares the complete sample of well-known bulge clusters, plus
the new sample, to clusters from
the dwarf galaxies Sagittarius and Canis Majoris 
listed in \citet{forbes10}, and from Gaia-Enceladus-Sausage listed in \citet{limberg22}.


Figure \ref{age-met} gives a metallicity versus age for the new clusters that have these values available,
excluding seven clusters that have no measurement of proper motion and/or radial velocity.
They are compared with the \cite{bica16} sample, and with the Sagittarius
and Canis Majoris clusters identified in \citet{forbes10}, and updates
 from \citet{kruijssen19}, plus Gaia-Enceladus-Sausage clusters and
the age-metallicity curve from \citet{limberg22}.
This figure shows that all known in situ  bulge clusters are older than  $> 11.5$\,Gyr,
and do not follow an AMR,  as is shown by the solid line
at 12.5 Gyr. 
This is the case
of a fraction of the new clusters, whereas the other clusters follow
the AMR of the ex situ clusters from the dwarf galaxies.
Although the age-metallicity curves of different dwarf galaxies are somewhat different, as can be seen for example in Figure 9 of  \citet{callingham22}, they are sufficiently close for us to plot
that relation from Gaia-Enceladus-Sausage.

For a deeper study, better derivations of ages with deeper photometric data are needed.
We also note that for AMRs it is important to have  relative ages, 
as well as homogeneous derivations of ages, as
is discussed by, for example, \citet{2011ApJ...738...74D}, \citet{vandenberg13}, and \citet{leaman13}.
We recognize that there are likely systematic differences between the different studies that are at least as large as the assigned random errors.
However, a homogenization of ages cannot be found in the literature
for either the well-known bulge clusters or the newly identified clusters.

Finally, radial velocity measurements are required to study the cluster orbits. Therefore, no firm
conclusion can be drawn yet for the clusters with ages estimated to be lower than 10\,Gyr,
but in principle they are compatible with an ex situ origin.

\begin{figure*}
\centering
\includegraphics[width=0.48\textwidth]{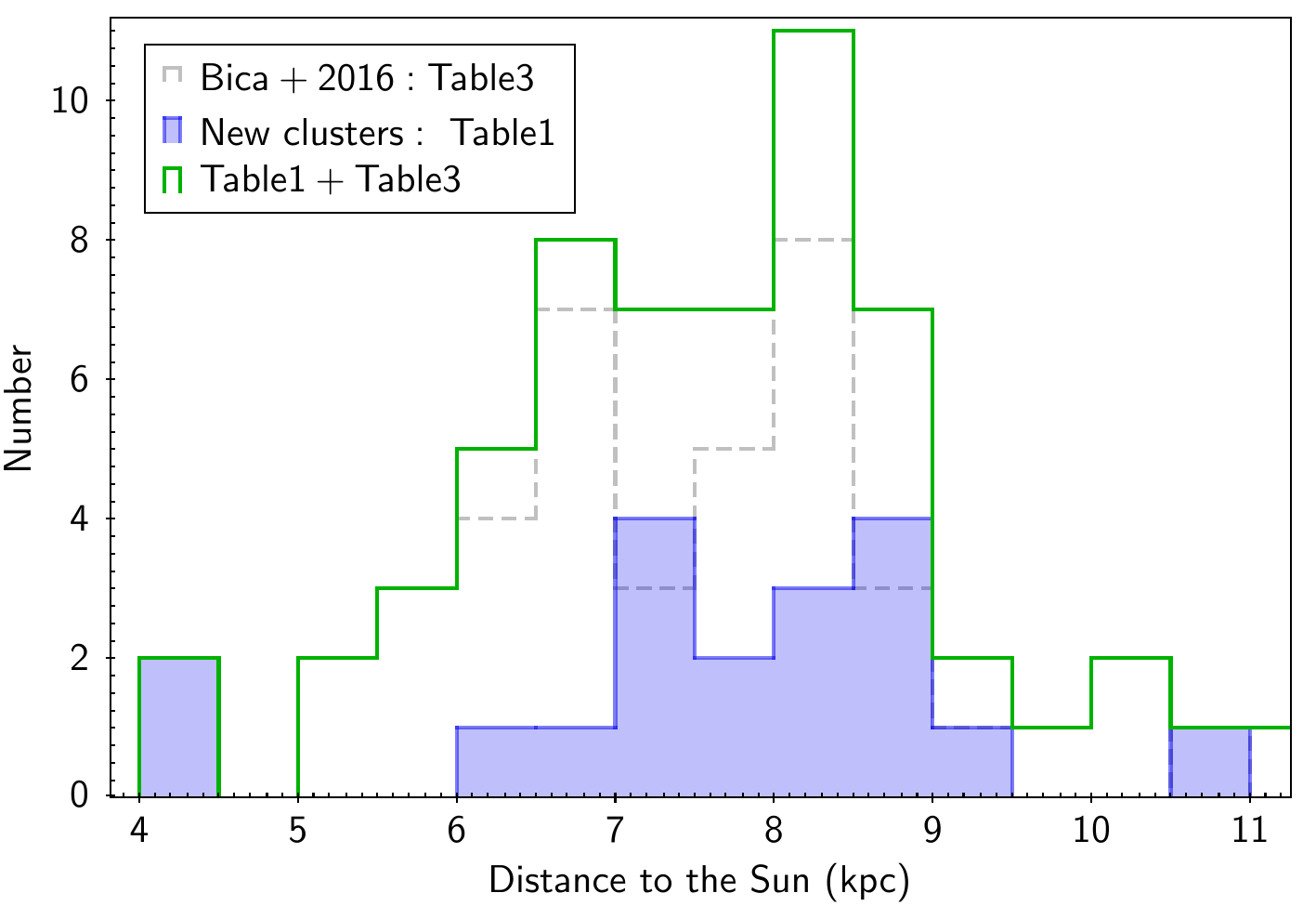}
\includegraphics[width=0.48\textwidth]{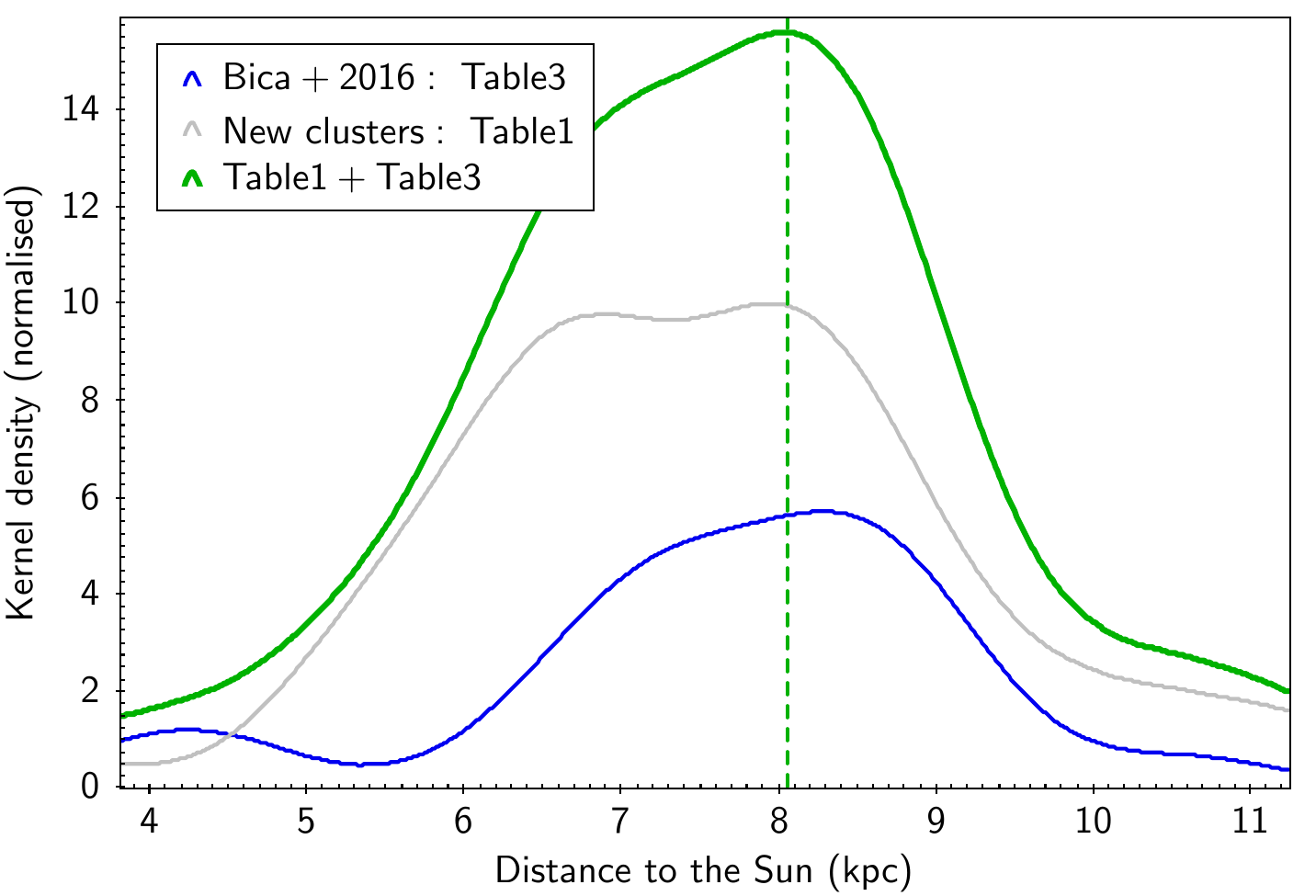}
\caption{Distribution of distances of the same cluster sample as in Figure~\ref{metallicity}.
Despite having the same smoothing, the kernel density estimations are noisier than the metallicity ones, but show that the \citet{bica16} and the full samples share a peak around $8.0-8.1$\,kpc, as was expected.
}
\label{distance} 
\end{figure*}

\subsection{Distances}

Figure~\ref{distance} shows the histogram
and distribution of distances to the Sun for the new bulge
clusters (Table~\ref{logbook}) compared with the compiled distances from \citet[Table~\ref{bica16}]{bica16}
and with the full cluster sample. The samples are the same as in Figure~\ref{metallicity} except for the cluster Glimpse02, which has no available distance in the literature.

The histogram
of new clusters shows two peaks at 7.2 and 8.5\,kpc, but the kernel density estimation shows them as a smooth distribution centered around $8.0$\,kpc.
This configuration of the histogram remains for different distance bins.
The full cluster sample preserves the distance distribution from \citet{bica16},
with a peak around $8.0-8.1$\,kpc and a tail pointing to smaller distances.
The peak is compatible, within 0.2\,kpc, with the Galactic Black Hole distance
of d$_{GCenter}$ =  $8.178\pm0.013_{\rm stat.}\pm0.022_{\rm sys.}$\,kpc \citep{abuter19}.
More recently, \citet{arakelyan18} have found 8.2\,kpc using GCs and dwarf galaxies, but although this might appear obvious now, it was not the case with the data available at the time of our estimation of d$_{GCenter} = 7.5$\,kpc \citet{bica06}.
\citet{bland-hawthorn16} and \citet{griv2023} provide a historical review of this value.

There is another issue regarding distances, which is the selective-to-total absorption $R_{\rm V}$ value.
While the subject is debated, the recent papers by \citet{nataf13}, \citet{nataf16}, and \citet{nataf21} lean toward a low reddening slope with $R_{\rm V}=2.5$. \citet{pottasch13}, on the other hand, oppose this conclusion \citep[see][]{stenborg16},
based on the analysis of spectra of planetary nebulae in the Galactic bulge (at $b\sim 3-6^{\circ}$).
These authors conclude with a strong statement: ``The suggestion that $R_{\rm V}$ is different in the Galactic bulge is incorrect. The reasons for this are discussed.'' 
A reddening problem is also suggested by \citet{vasiliev21},
comparing \textit{Gaia} Early DR3 parallaxes with optical distances of bulge clusters. 
The inverted parallaxes give longer distances than the optical ones.

Some questions remain, however, such as why we should have such an anomalous reddening law. There is no evidence around the Sun of a reddening law with $R_{\rm V} < 3$.
We may eventually have much higher values.

Another question is whether we have an additional problem of a limiting distance, or a peculiar cloud, between us and the bulge, and in that case where it is.
\citet{nataf21} discusses a peculiar ``great dark lane'' when citing \citet{minniti14}, but there is no theoretical support for this, and it would need very large dust grains.
The point is not negligible, because it can also introduce a bias in the age measurements with the isochrones, inducing a trend of age versus Galactocentric distance.

The comparison of the \textit{Gaia} Early DR3 distances by 
\citet[hereafter BV21]{baumgardt21}
with those
given in the literature for the new clusters, however, does not help because
of the uncertainties in the parallaxes.

\section{Conclusions}
\label{sec:5}

We compiled a sample of 39 new GCs and two candidates, mostly in the Galactic bulge, which have been identified since the 
\citet[2010 edition]{harris10} catalog and \citet{bica16}.
The number of confirmed bulge GCs in the Galaxy has considerably increased, with the
difference being that the new clusters have very low masses.
The new full sample of bulge clusters shows a metallicity distribution with peaks at
$\rm{[Fe/H]} = -1.08\pm0.35$ and $-0.51\pm0.25$\,dex.

The age-metallicity plot shows two different aspects: all well-known bulge clusters are older
than 11.5 Gyr, and 13 of the new clusters are in this category; that is, they are all old and
no AMR is seen, as has been shown by studies in the literature, such as those by
\citet{forbes10}, \citet{forbes20}, \citet{kruijssen20}, and \citet{callingham22}.

For the 11 clusters younger than 11.5 Gyr, the AMR appears to be compatible with that of the ex situ
dwarf galaxies Sagittarius, Canis Majoris, and Gaia-Enceladus-Sausage. 

Finally, we conclude that it is likely that
very few, if any, other massive GCs in the Galactic bulge are still to be found, although we cannot exclude that massive clusters with low densities remain undetected. Given the dense environment, it is expected that some of the lowest mass clusters could also be dispersing, and thus be missed in the dense bulge field.

A next step in the study of these clusters should be spectroscopic analyses, in particular to derive their radial velocities. This will allow us to derive their orbits and to better understand their origin.

\begin{acknowledgements}
  We are grateful to the referee for very helpful suggestions,
  and to Stefano O. Souza for helpful comments.
EB and BB acknowledge partial financial support
from CAPES - Finance code 001, CNPq and FAPESP.
SO acknowledges support from PRIN MIUR2022 Progetto "CHRONOS" (PI: S. Cassisi) financed by the
European Union - Next Generation EU,  the support of the University of Padova, DOR  Piotto 2022, Italy.
RAPO acknowledges the FAPESP PhD fellowship 2018/22181-0.
\end{acknowledgements}


\bibliographystyle{aa} 
\bibliography{bibliog.bib}

\begin{appendix}

\end{appendix}
\end{document}